\documentclass[12pt]{article}

\begin {document}
\begin{flushright}
{\small
SLAC--PUB--8587\\
August 2000\\}
\end{flushright}

\begin{center}
{{\bf\LARGE
Conventional Beams or Neutrino\\[.20ex] Factories: 
The Next Generation of \\[1ex] Accelerator-Based Neutrino Experiments
}\footnote{Work supported by
Department of Energy contract  DE--AC03--76SF00515.}}

\bigskip \bigskip
Burton Richter\\
Stanford Linear Accelerator Center\\
Stanford, California  94309-0450 \\
\end{center}

\vfill

\begin{center}
{\bf\large
Abstract}
\end{center}

The purpose of this paper is to provoke a discussion about the right
next step in accelerator-based neutrino physics.  In the next five years
many experiments will be done to determine the neutrino mixing
parameters.  However, the small parameters 
$\theta_{13},\ \Delta m_{21}^2$, and the CP
violating phase are unlikely to be well determined.  Here, I look at the
potential of high-intensity, low-energy, narrow-band conventional
neutrino beams to determine these parameters.  I find, after roughly
estimating the possible intensity and purity of conventional neutrino
and anti-neutrino beam, that $\sin^2\theta_{13}$ can be measured if greater than a
few parts in ten thousand, $\Delta m_{21}^2$ can be measured if it is greater than
$4\times 10^{-5}$ (eV)$^2$, and the CP violating phase can be measured if it is
greater than 20$^\circ$ and the other parameters are not at their lower bounds.
If these conclusions stand up to more detailed analysis, these
experiments can be done long before a muon storage ring source could be
built, and at much less cost.

\vfill
\newpage

\normalsize

Neutrino physics has become one of the central studies of current
high-energy physics.  Experiments on neutrinos from the sun have shown
that not enough of them reach the earth to be consistent with the sun's
energy output.  Experiments on neutrinos produced in the atmosphere have
shown that muon-type neutrinos seem to be depleted while passing through
the earth compared to those incident on detectors directly from above.
Neutrinos no longer seem to be the zero-mass passive participants in the
sub-nuclear world that they were thought to be only a few years ago.
Instead, the picture that is emerging shows a system of three, or
possibly four, neutrino-mass eigenstates that mix in different amounts
to form the familiar flavor eigenstates, the electron, muon and tau
neutrino types.  What is produced in a reaction is a flavor eigenstate
which evolves as it travels because of a change in relative phases of
the different mass eigenstates arising from their relative mass
differences.

Each year brings new information that is beginning to fill in a still
incomplete picture.  The continuation of several current experiments and
a host of soon to be begun experiments will further fill in the picture.
However, enough is known now to indicate that certain of the interesting
parameters, in particular the mixing angle known as $\theta_{13}$, the mass
difference $\Delta m_{21}^2$, the magnitude if any of CP violation in the neutrino
sector, and the ordering of the neutrino masses cannot be well
determined, if determined at all, from the current crop of running and
planned experiments.  Thus attention has begun to turn toward
high-intensity accelerator sources to get further information.

Much of this attention has been focused on the potential of high-energy
(20 GeV and above) muon storage rings as candidate sources.
\cite{ref:1}  These
devices require considerable R\&D to determine whether they are indeed
feasible, and, if feasible, will be very expensive.  In this paper I
will discuss the potential of high-intensity, conventional neutrino
beams to get at these small parameters.  My conclusion is that
low-energy, conventional muon-neutrino beams of attainable intensity are
serious candidates for the next generation systems.

Before analyzing what can be done in the real world, it is useful to
look at a primitive, two-neutrino, model to illustrate why low energy
may be better than high energy in untangling the puzzle.  Consider two
species with a small mixing amplitude between them.  The
``signal-to-noise'' ratio in an experiment looking for the appearance of
neutrino type two in a beam of neutrino type one is given by 
\begin{equation}
\frac{P(\nu_1\rightarrow \nu_2)}{P(\nu_1\rightarrow \nu_1)} =
\frac{A^2\sin^2(\Delta m^2L/4E)}{1-A^2\sin^2(\Delta m^2L/4E)}
\end{equation}
 where $A$ is the mixing amplitude, $\Delta m^2$ is the
difference of the squares of the masses, $L$ is the distance from the
source to the detector, and $E$ is the beam energy.  The optimum
signal-to-noise ratio comes when the sine term is equal to one, {\em i.e.},
$\Delta m^2L/4E$ is equal to an odd integer multiple of $\pi/2$.  However, all of the
muon storage ring designs have high energy, making this factor small
with the known mass difference.  Hence, a very sophisticated background
rejection is required in the detector.

It is also sometimes argued that, because neutrino cross sections
increase with energy and the flux of neutrinos increases with the square
of the energy of the parent particle, high-energy beams are better than
low energy ones.  However, in looking for the small terms that are
unlikely to be determined by the present round of experiments, the
appearance of a neutrino species different from the primary species
gives the most sensitive tests.  This probability is proportional to
$E^{-2}$, leaving only a single power of the energy as a potential advantage
for the high-energy beams.  Thus, this rationale for the choice of high
energy is not as overwhelming as it might appear.

In addition, in the real three-neutrino world, there is a further
complication from high-energy beams, tau-lepton production.  In the
scenarios using a storage-ring source, tau-lepton production is a
potentially serious complication in determining the small parameters,
since its leptonic and semi-leptonic decays can make a tau event look
like an electron or a muon event.  It is proposed to solve this problem
by determining, for example, the sign of electrons produced by muon
neutrinos \cite{ref:2} which would be effective but is very difficult in a large
detector.  It is a simplification if one can stay below the tau-lepton
production threshold.

Now I turn to the standard three-neutrino formulation (recent results
from Super Kamiokande appear to give 95\%\ confidence in this
formulation).  Equation (3) includes the small terms that are often
dropped in oscillation analyses.  In Eq.  (3), $C_{ij}$ means 
$\cos\theta_{ij},\,  S_{ij}$
means $\sin\theta_{ij},\, \Delta m_{ij}^2$  means 
$m_i^2 - m_j^2$, $\delta$ is the CP violating phase, $L$ is
the distance from the source to the detector, and $E$ is the neutrino
energy.  The matter effect is given by
\begin{equation} 
a = 2\sqrt 2 \, G_Fn_eE = 7.6 \times 10^{-5}\, \rho(gm/cm^3)\, E(GeV) \qquad
(eV)^2
\end{equation}
 where $G_F$ is the Fermi
weak-coupling constant, and $n_e$ is the electron density.  Equation (3) is
exact when the matter term of Eq.  (2) is zero; otherwise $a/\Delta m_{31}^2$ is
assumed to be small.\cite{ref:3}
\begin{eqnarray}
&& P(\nu_\mu\rightarrow \nu_e) = 4C_{13}^2 S_{13}^2S_{23}^2\sin^2
\frac{\Delta m_{31}^2L}{4E} \times 
\left(1+
\frac{2a}{\Delta m_{31}^2}\
\left(1-2S_{13}^2\right)\right) \nonumber \\
&&+ 8C_{13}^2S_{12}S_{13}S_{23}(C_{12}C_{23}\cos\delta-S_{12}S_{13}
S_{23})\cos 
\frac{\Delta m_{32}^2L}{4E}\sin 
\frac{\Delta m_{31}^2L}{4E} \sin 
\frac{\Delta m_{21}^2L}{4E} \nonumber \\
&&  -8C_{13}^2 S_{13}^2 S_{23}^2 \cos 
\frac{\Delta m_{32}^2 L}{4E}\sin 
\frac{\Delta m_{31}^2L}{4E}\, 
\frac{aL}{4E}\,
\left(1-2S_{13}^2\right) \\
&& - 8C_{13}^2 C_{12} C_{23} S_{12} S_{13} S_{23} \sin\delta \sin
\frac{\Delta m_{32}^2L}{4E} \sin 
\frac{\Delta m_{31}^2L}{4E} \sin
\frac{\Delta m_{21}^2L}{4E}  \nonumber \\
&& + 4S_{12}^2C_{13}^2 \left\{ C_{12}^2 C_{23}^2 
       + S_{12}^2S_{23}^2S_{13}^2 - 2C_{12} C_{23} 
S_{12} S_{23} S_{13}\cos \delta\right\} \sin^2
\frac{\Delta m_{21}^2L}{4E} \ .\nonumber
\end{eqnarray}
 It is not my
purpose here to do an exhaustive analysis of all of the possibilities,
but to show some examples of the potential, and in what follows I will
assume that $\theta_{23} = \theta_{12} = \pi/4$.  The Super Kamiokande 
atmospheric data
favors this for $\theta_{23}$ and their solar data favor the large mixing-angle
solution for the solar neutrino deficit.  I will also take $\Delta m_{32}^2 = 
3 \times 10^{-3}\, (eV)^2$ which is the central value of the latest Super-K data,
\cite{ref:4} $\Delta m_{31}^2 \approx \Delta m_{32}^2$,
$ S_{13}^2 \ll 1$, and the matter term $(a) = 0$.

Now, look at the results of four experiments; $\nu_\mu \rightarrow \nu_e$ and 
$\bar\nu_\mu \rightarrow \bar \nu_e$ at two values of $L/E$ such that 
$\Delta m_{32}^2 L/4E$ equals $\pi/2$ and
$\pi$. The result when the $\Delta m_{32}^2$ term is $\pi/2$ is, 
\begin{equation}
P_+(\pi/2) = P(\nu_\mu\rightarrow \nu_e) + P(\bar \nu_\mu \rightarrow \bar \nu_e)
= 4S_{13}^2+\sin^2 \frac{\Delta m_{21}^2L}{4E} 
\end{equation}
\begin{equation}
P_-(\pi/2) = P(\nu_\mu\rightarrow \nu_e) - P(\bar \nu_\mu \rightarrow \bar \nu_e)
= -4S_{13} \sin\delta \sin \frac{\Delta m_{21}^2L}{4E} 
\end{equation}
and, when the mass term is $\pi$, is 
\begin{equation}
 P_+(\pi) = P(\nu_\mu\rightarrow \nu_e) + P(\bar \nu_\mu\rightarrow \bar \nu_e)
= \sin^2 \frac{\Delta m_{21}^2L}{4E}
\end{equation}
\begin{equation}
P_-(\pi) = P(\nu_\mu\rightarrow \nu_e) - P(\bar \nu_\mu\rightarrow \bar \nu_e) = 0 
\ .
\end{equation}
The next question is, are there enough neutrinos to do useful
experiments.  It will require a detailed design of a beam to get a
precise answer, but we can get a rough idea by extrapolating from work
already done at FNAL on the MINOS beam design and from the FNAL storage
ring source study.\cite{ref:1} The present FNAL proton beam power for the MINOS
beam is about 400~ kW.  The beam intensity is limited by space-charge at
injection into the 8-GeV Booster and at injection into the 100--GeV Main
Injector.  If the FNAL linac's energy was increased to about 1 GeV and
the Booster's energy increased to about 16 GeV, the FNAL beam on target
could be increased to about 4 MW. \cite{ref:5} The improvements to the injector
chain are straightforward.  The proton target is more difficult than the
present system and will certainly require more shielding.  Target R\&D is
being intensively pursued for the Spallation Neutron Source and should
present no fundamental problems for neutrino production.  The target for
a conventional beam is somewhat easier per unit beam power than that
for a muon storage ring source.  The conventional beam is designed to
produce pions of multi-GeV energy, while the muon source uses pions of
low energy.  For maximum production, the conventional source is thus
thinner and absorbs less power than the storage ring source.

The comparative yields from a high-intensity, conventional, wide-band
beam and a muon storage ring can be estimated by scaling the numbers
given in the FNAL report.  Table 1 shows the numbers.  Using those
numbers, a 20 kiloton detector at 732~km would see about $9 \times 10^4$ events
per year from the low-energy, wide-band, conventional beam.

\begin{table}[h]
\begin{center}
\caption[]{Muon-neutrino charged-current events per kiloton year, assuming no
oscillation, for a 4--MW upgraded FNAL wide-band conventional source, and
for a 20-GeV muon storage ring giving 10$^{20}$ decays per year.  The numbers
are scaled from Ref. [1].}\bigskip
\begin{tabular}{|l|c|c|c|}
\hline
Source  Type &Mean $\nu$ Energy (GeV) &Baseline (km)& N \\[-1ex]
& & & CC-events/kT-year \\ \hline \hline
Conventional & 3 & 732 & 4.6 $\times$ 10$^3$ \\
Conventional & 6 & 732 & 1.4 $\times$ 10$^4$ \\
Conventional & 12 & 732 & 3.2 $\times$ 10$^4$ \\ \hline
Storage Ring & 15 & 732 & 1.2 $\times$ 10$^4$ \\ \hline
\end{tabular}
\end{center}
\end{table}

At 732~km, the neutrino beam energies required to have $(\Delta m_{32}^2 L/4E)$
equal to $\pi/2$ and $\pi$ would be about 2 GeV and 1 GeV respectively ($\pi$ mesons
of about 9 and 4.5 GeV).  There is no design available for a narrow band
beam, and so I use the spectrum of the MINOS low-energy beam to make a
rough estimate.  The spectrum has a low-energy peak and a high-energy
tail, each generating about equal number of events \cite{ref:6} Much improved
focusing can be achieved in a narrow band beam, which increases the
flux.  The peak in the MINOS spectrum has a width of about $\pm$ 30\%, while
it is desirable for background rejection to have a narrower width, thus
reducing the flux.  I will assume that the spectrum has a width of
($\pm$ 10--20\% and that the focusing and width effects balance out in
determining the beam intensity.  Finally, I will give up another factor
of two because this estimate may be too optimistic, or the FNAL Main
Injector upgrade may not be capable of reaching the 4 MW power level.
The result for neutrinos in the 20-kiloton detector at 732~km from the
source is
\begin{equation}
Y_{cc}(2\ GeV) = 2.5 \times 10^4 \qquad {\rm   events\ per\ year.} 
\end{equation}
A 1--GeV beam will have a lower rate.  There will be more pions in the
decay channel, but the neutrino beam divergence will increase by a
factor of two and the charged-current neutrino cross-section will
decrease by a factor of two resulting in a reduction of about a factor
of four in event rate
\begin{equation}
Y_{cc} (1\ GeV) = 6\times 10^3 \qquad {\rm    events\ per\ year}.  
\end{equation}
Note that antineutrino
cross-sections are about half of those for neutrinos.

Electron neutrinos in the beam come from the chain $\pi \rightarrow \mu \rightarrow e$
 in the decay channel.  They amount to about 0.1\%\ total at the far detector, and
amount to $(1-2) \times 10^{-4}$  in the muon neutrino energy range.  However, a
more serious background will be neutral-current interactions initiated
by muon and tau neutrinos in the detector.  These can give $\pi^0$ mesons,
and highly asymmetric conversions of the $\pi^0$ decay $y$'s can
be confused with electron-neutrino-charged current events.\cite{ref:7} These have
to be controlled by detector design.  Here I will assume that they give
a background of 0.1\%\ of the ``unoscillated''  muon-neutrino flux.

The optimum data-taking strategy depends on ``a priori'' knowledge.  If,
for example, KamLAND determines $\Delta m_{21}^2$, there is no reason to take data
at 1 GeV.  Here, I will simply assume that data equivalent to 10$^4$
charged-current events (in the absence of oscillation) is collected for
$\nu_\mu$ and $\bar \nu_\mu$ at both 2 GeV and 1
GeV.  I also ignore the effect of the finite energy spread in the beam.

From Eq.  (6), the minimum detectable value of $\Delta m_{21}^2$ different from
zero at the three standard-deviation level is 
\begin{equation}
\left(\Delta m_{21}^2\right)_{\rm min} = 4.2 \times 10^{-5}\ (eV)^2 \ .
\end{equation}

This is independent of $\theta_{13}$ as long as $C_{13}$ is near one.  It is near the
lower end of the solar Large Mixing Angle solution allowed region and,
if this large, might be measurable by the KamLAND experiment.

Equation (4) couples $\Delta m_{21}^2$ and $S_{13}^2$.  
Table 2 gives the 3$\sigma$ lower bound
for $S_{13}^2$ for various values of $\Delta m_{21}^2$ ranging from the minimum detectable
in this proposed experiment to the maximum in the solar LMA solution.
What is happening, as the usually ignored $\Delta m_{21}^2$ term increases, is that
more events have to come from the $S_{13}^2$ term to meet the three standard
deviation requirement.

\begin{table}[h]
\begin{center}
\caption[]{Three standard deviation lower bound on the determination of 
$S_{13}^2$ for various values of $\Delta m_{21}^2$.}
\medskip
\begin{tabular}{|c|c|}
\hline
$\Delta m_{21}^2\ (eV)^2$ & $S_{13}^2$ \\[1ex]
\hline
$4 \times 10^{-5}$ & $5\times 10^{-4}$ \\[1ex]
$2 \times 10^{-4}$ & $9 \times 10^{-4}$ \\
$1 \times 10^{-3}$ & $4 \times 10^{-3}$ \\
\hline
\end{tabular}
\end{center}
\end{table}

The last item to look at is the sensitivity to the CP violating phase $\delta$.
This is given in Eq. (5) and depends on both 
$S_{13}$  and $\Delta m_{21}^2$ (note also
that the sign of $\Delta m_{21}^2$ comes in).  If either of these is very small, CP
violation becomes impossible to measure in this or any other experiment.
To give an idea of the sensitivity, I will take $\Delta m_{21}^2 = 2 \times 10^{-4}$, the
center of its range, and take $S_{13}^2$ at the 3$\sigma$ limit for this $\Delta m_{21}^2$,
{\em  i.e.} $ 9 \times 10^{-4}$.  The minimum detectable value for the phase that 
differs from zero by three $\sigma$ is, 	
\begin{equation}
| \sin \delta|_{\rm min} = 0.35
\end{equation}		
where Eq.  (4) has been used to determine the error on the non-CP
violating yield.

The determination of $\delta$ is most sensitive to the matter term.  With
Sato's \cite{ref:3} first order analysis, the matter term is 30\%\ of the CP
violating term for the parameters used above.  I believe the matter
effect needs to be taken to higher order.  It is interesting to note
that as the neutrino beam energy is increased, the matter term gets
bigger while the CP violating term gets smaller, another argument for
low-energy beams.

In summary, I have analyzed here the potential of high-intensity,
low-energy, narrow-band conventional neutrino beams.  The ``gedanken''
experiments outlined here give interesting limits on the measurement of
the small terms among the neutrino-mixing parameters.  These limits are
better than those from the "entry level" storage ring neutrino
factory. \cite{ref:8} It is well worth the time of the experts to see if my
assumptions on potential beam intensity and purity, and background
rejection are reasonable.  If they are, these experiments can be carried
out sooner, and at less cost than those with a muon storage ring source.


\begin{thebibliography}{99}

\bibitem{ref:1}
Fermilab-FN-692, May 10, 2000.

\bibitem{ref:2}
{\em Op cit.} FNAL report.

\bibitem{ref:3}
I thank Jo\~ ao Silva of Instituto Superior de Engenharia de Lisboa,
Portugal, for showing me the exact expression in the absence of matter
effect.  The first order matter effects come from J. Sato,
hep-ph/0006127, 13 June 2000.  In determining the small parameters it
may not be consistent to keep first order terms in the matter effect
while keeping all of the other terms.  In working out examples I have
set the matter effect equal to zero.  It will eventually have to be put
in consistently.

\bibitem{ref:4}
  The latest Super Kamiokande data is not yet published,
but is available in the talks of Y. Takeuchi and T. Toshito in session
PA-086 at the XXX International Conference on High-Energy Physics,
Osaka, July 2000,  {\tt http://ichep2000.hep.sci.osaka-u.ac.jp/}.

\bibitem{ref:5}
  There are many other things
that would need doing, but the Linac and Booster work are the costly
ones.

\bibitem{ref:6}
 MINOS Technical Design Report, Figure 3.3, \hfill\break
{\tt  http://www.hep.anl.gov/ndk/hypertext/minos$\underline{}$tdr.html}. 
The figure
shows the interaction energy spectrum for the low, medium, and high
energy beams.  The figure also shows the limit for ``perfect focusing'' of
all energies.  This limit is much easier to approach in a narrow-band
beam than in a wide-band beam.

\bibitem{ref:7}
 This was the limiting factor in BNL
Experiment 776 [see Phys. Rev. Lett.  {\bf 62}, 2237 (1989)] .

\bibitem{ref:8}
 See, for example,
V. Barger {\em et al.}, hep-ph/0003184 v2, July 2000.

\end{thebibliography}
\end{document}